\documentclass[aps,prl,groupedaddress,showpacs,notitlepage]{revtex4-1}
\usepackage{graphicx}
\usepackage{rotating}
\usepackage{amsmath}
\usepackage{amssymb}
\usepackage{mathptmx}
\usepackage{subfigure}

\begin{document}

\newcommand{\hdblarrow}{H\makebox[0.9ex][l]{$\downdownarrows$}-}
\newcommand{\I}  [0] {{\rm i}}            
\newcommand{ \ket}[1] {\left| #1 \right\rangle}
\newcommand{ \bra}[1] {\left\langle #1 \right|}
\newcommand{ \rvec}[0] {{\mathbf r}}
\newcommand{ \pvec}[0] {{\mathbf p}}
\newcommand{ \kvec}[0] {{\mathbf k}}
\newcommand{ \qvec}[0] {{\mathbf q}}

\def\hm#1{\frac{\hbar^2}{#1m}}
\def\he#1{$^{#1}$He}

\def\comment#1#2{\bigskip\hrule\smallskip\noindent{\bf#1 :\ }#2
\smallskip\hrule\bigskip}

\def\Im{{\cal I}m}
\def\Re{{\cal R}e}

\title{Dynamic many-body theory:\\
Dynamic structure factor of two-dimensional liquid $^4$He}

\author{E. Krotscheck$^{1,2}$ and T. Lichtenegger$^{1,2}$}

\affiliation{$^1$Department of Physics, University at Buffalo, SUNY
Buffalo NY 14260}
\affiliation{$^2$Institut f\"ur Theoretische Physik, Johannes
Kepler Universit\"at, A 4040 Linz, Austria}

\date{\today}

\begin{abstract}
  We calculate the dynamic structure function of two-dimensional
  liquid $^4$He at zero temperature employing a quantitative
  multi-particle fluctuations approach up to infinite order. We
  observe a behavior that is qualitatively similar to the
  phonon-maxon-roton-curve in 3D, including a Pitaevskii plateau
  (L. P. Pitaevskii, Sov. Phys. JETP {\bf 9}, 830 (1959)).  Slightly
  below the liquid-solid phase transition, a second weak roton-like
  excitation evolves below the plateau.
\end{abstract}



\maketitle

\section{Introduction}

The static and dynamic structure of few-layer films of liquid helium
absorbed on solid substrates at low temperatures has been studied
experimentally, {\em e.g.}\ within neutron scattering
measurements~\cite{lambert1977surface,thomlinson-80,LauterExeter,LauterJLTP},
and theoretically~\cite{ChangCohen75,Surface1,Surface2,JiWortis,%
  OldSurf,Ceejay,Gernoth2,filmexc}. The earliest investigations of
excitations \cite{ChangCohen75,Surface1,Surface2,JiWortis,Gernoth2}
were based on generalizations of Feynman's theory of excitations in
the bulk liquid~\cite{Feynman} and therefore only qualitative. Later
work \cite{Ceejay,filmexc,ApajaKroHectorite2} employed correlated
basis functions (CBF) theory \cite{JaFe,JaFe2,Chuckphonon}.  These
methods are simple enough for the application to non-uniform
geometries including the inhomogeneity of the substrates and the
non-trivial density profile of the films. Agreement with measurements
of the dynamic structure was either semi-quantitative, or required
some phenomenological input for a quantitative description of the
various excitation types seen in the experiments~\cite{filmexpt} such
as ``layer-phonons'', ``layer-rotons'', or ``ripplons''.

Since then the development of theoretical tools for describing the
dynamics of bulk quantum liquids has made significant progress,
providing a quantitative description in the experimentally accessible
density range for low and intermediate momenta, probing the
short-range structure of the
system~\cite{eomIII,He4ModeMode,2p2h,SDFHe3}. Due to the increasingly
complicated form of more elaborate methods, application to
inhomogeneous geometries is less straightforward. Building on the
success of our method for both bulk \he4 \cite{QFS09_He4} and \he3
\cite{2p2h,SDFHe3,Nature_2p2h}, we here investigate mono-layer films
of $^4$He which can be treated as strictly two-dimensional liquids.

Recently, novel numerical methods
\cite{PhysRevB.82.174510,PhysRevB.88.094302,NGMV2013} have appeared
that give access to dynamic properties of quantum fluids.  These are
algorithmically very important developments that will ultimately aide
in the demanding elimination of background and multiple-scattering
events from the raw data. However, it is generally agreed upon that
the model of static pair potentials like the Aziz interaction
describes the helium liquids accurately. Hence, given sufficiently
elaborate algorithms and sufficient computing power, such calculations
must reproduce the experimental data.  The aim of our work is somewhat
different: The identification of physical effects like phonon-phonon,
phonon-roton, roton-roton, maxon-roton \dots\ couplings that lead to
observable features in the dynamic structure function is, from
simulation data, only possible {\em a-posteriori\/} whereas the
semi-analytic methods pursued here permit a direct identification of
these effects, their physical mechanisms, and their relationship to
the ground state structure directly from the theory.

\section{Theoretical framework}

The behavior of $N$ identical, non-relativistic particles in an
external field $U_{\rm ext}(\rvec)$, interacting via a pair
potential $V_{\rm int}(\rvec,\rvec')$, is governed by a microscopic
Hamiltonian
\begin{equation}
 H_0=-\sum_{i=1}^N \frac{\hbar^2}{2m}\nabla_i^2+\sum_{i=1}^N
U_{\rm ext}(\rvec_i)+\sum_{\substack{i,j=1 \\ i<j}}^N V_{\rm int}(\rvec_i,\rvec_j).
\end{equation}
The ground state is written in the Feenberg form \cite{FeenbergBook}
\begin{equation}
 |\Psi_0\rangle=e^{\frac{1}{2}U}|\phi_0\rangle,
\label{eq:wavefunction}
\end{equation}
where $|\phi_0\rangle$ is a non- or weakly-interacting model wave
function containing the appropriate symmetry and statistics of the
system, and
\begin{equation}
 U(\{\rvec_k\})=\sum_{i=1}^N u_1(\rvec_i)
+\sum_{\substack{i,j=1 \\ i<j}}^N u_2(\rvec_i,\rvec_j)
+\sum_{\substack{i,j,k=1 \\ i<j<k}}^N u_3(\rvec_i,\rvec_j,\rvec_k)+\dots
\label{eq:correlations}
\end{equation}
is the correlation operator consisting of $n$-particle correlation
functions $u_n$.

For homogeneous Bose systems such as three- and two-dimensional
$^4$He, $|\phi_0\rangle$ can be chosen to be 1 and the wave function
(\ref{eq:wavefunction},\ref{eq:correlations}) is in principle exact.
The empirical Aziz potential \cite{Aziz} as interaction between the
helium atoms has turned out to lead to results in quantitative
agreement with experiments, see Ref.~\onlinecite{JordiQFSBook} for a
review. With minimal phenomenological input, the same accuracy can be
obtained with integral equation methods \cite{ChaC,EKthree}.  In that
case, the correlation functions are optimized by minimizing the ground
state energy $E_0$, viz.
\begin{equation}
 \frac{\delta E_0}{\delta u_n}=\frac{\delta }{\delta u_n}
\frac{\langle \Psi_0|H_0|\Psi_0\rangle}{\langle \Psi_0|\Psi_0\rangle}=0.
 \label{eq:eula}
\end{equation}

Dynamics is treated along basically the same lines.
In the presence of a time-dependent external perturbation 
\begin{equation}
 \delta H(\{\rvec_k\};t) = \sum_i \delta U_{\rm ext}({\bf r}_i;t),
\end{equation}
the time-dependent generalization of the ground state wave function
(\ref{eq:wavefunction}) is
     \begin{equation}
\left|\Psi(t)\right\rangle = \frac{e^{-\I E_0 t/\hbar} \,
        e^{\frac{1}{2}\delta U(t)}\left|\Psi_0\right\rangle}{
        \left[\left\langle\Psi_0\left|e^{\Re \delta U(t)}
        \right|\Psi_0\right\rangle\right]^{1/2}},
\label{eq:excwave}
\end{equation}   
where
\begin{equation}
\delta U(\{\rvec_k\};t) = \sum_i\delta u_1({\bf r}_i;t) + \sum_{i<j}
                \delta u_2({\bf r}_i,{\bf r}_j;t)+\cdots
\label{eq:exop}
\end{equation}
is the complex {\it excitation operator\/}. Its components, the
fluctuations $\delta u_n({\bf r}_1,\ldots,{\bf r}_n;t)$ of the
correlation functions, are determined by the least action principle
\cite{KramerSaraceno,KermanKoonin}
\begin{equation}
\delta \int dt \left\langle\Psi(t)\left|H_0+\delta H(t)
        -\I\hbar\frac{\partial}{\partial t}
        \right|\Psi(t)\right\rangle = 0,
\label{eq:action}
\end{equation}
which generalizes the Euler-Lagrange Eq.~\eqref{eq:eula} to the
time-dependent case.

\section{Multi-particle fluctuations and density-density
response}
For weak external perturbations, the relationship between the perturbing
external field and the
induced density fluctuation
\begin{equation}
 \delta \rho(\rvec;t)=\int d^3r'dt'
\rho_0(\rvec)\chi(\rvec,\rvec';t,t')\rho_0(\rvec')\delta U_{\rm ext}(\rvec';t')
+{\cal O}\big(\delta U_{\rm ext}^2\big)\label{eq:linres}
\end{equation}
is linear and defines the density-density response
function $\chi(\rvec,\rvec';t,t')$. In homogeneous, isotropic
geometries where the ground state density ${\rho_0(\rvec)=\rho_0}$ is
constant, the density-density response function is most conveniently
formulated in momentum and energy space and defines the
dynamic structure function
\begin{equation}
 S(k, \hbar\omega)=-\frac{1}{\pi}\Im\, \chi(k, \hbar\omega),
\end{equation}
spelled out here for zero temperature and consequently
$\hbar\omega>0$.

The truncation of the sum of many-particle fluctuations (\ref{eq:exop})
defines the level of our treatment of the dynamics.
For example, the single-particle approximation
\begin{equation}
 \delta U_{\rm F}(t) = \sum_i\delta u_1({\bf r}_i;t),
\end{equation}
for the fluctuations leads to the time-honored Feynman dispersion
relation \cite{Feynman}
\begin{equation}
 \varepsilon_{\rm F}(k)\equiv \frac{\hbar^2k^2}{2mS(k)}.
 \label{eq:feyn}
\end{equation}
Here, $S(k)$ is the static structure function which can be obtained
from experiments or ground state calculations.  In this approximation,
$S(k, \hbar\omega)$ is described by a single mode located at the
Feynman spectrum.

The importance of including at least two-particle fluctuations $\delta
u_2(\rvec,\rvec';t)$ was first pointed out by Feynman and Cohen in
their seminal work on ``backflow'' correlations
\cite{FeynmanBackflow}. A somewhat more formal approach was taken by
Feenberg and collaborators who derived a Brillouin-Wigner perturbation
theory in a basis of correlated wave functions~\cite{JaFe,JaFe2,JacksonSkw,%
LeeLee,Chuckphonon}. These approaches
determine, rigorously speaking, only the energy of the lowest-lying
mode.  The equations of motion method \eqref{eq:action} employed here
provides access to the full density-density response function
\begin{equation}
 \chi(k,\hbar\omega)=\frac{S(k)}{\hbar\omega-\Sigma(k,\hbar\omega)+
\I\eta}+\frac{S(k)}{-\hbar\omega-\Sigma^*(k,-\hbar\omega)+\I\eta},
 \label{eq:cbf}
\end{equation}
where $\Sigma(k,\hbar\omega)$ is the phonon self-energy.  In
practically all applications, the excitation operator (\ref{eq:exop})
has been truncated at the two-body level and the convolution
approximation was used \cite{PPA2,ChaC} which is simple enough to be
employed in non-uniform geometries \cite{filmexc,filmexpt}. Then, the
self-energy has the form
 \begin{equation}
 \Sigma(k,\hbar\omega)=\varepsilon_{\rm F}(k) +
\frac{1}{2} \int \frac{{\rm d}^dp_1{\rm d}^dp_2}{(2\pi)^d\rho}
\delta(\textbf{k}+\textbf{p}_1+\textbf{p}_2)
\frac{|{V}^{(3)}(\kvec; \pvec_1,\pvec_2)|^2}
{\hbar\omega-\varepsilon_{\rm F}(p_1)-\varepsilon_{\rm F}(p_2)+\I\eta},
\label{eq:sigmacbf}
\end{equation}
where $d$ is the dimension of the system.  The
three-body vertex ${V}^{(3)}(\kvec; \pvec_1,\pvec_2)$ describes the
decay of a density fluctuation with wave vector $\kvec$ into two waves
with wave vectors $\pvec_1$ and $\pvec_2$. It can be calculated
in terms of ground state quantities, its general form
is \cite{eomI}
\begin{equation}
{V}^{(3)}(\kvec; \pvec_1,\pvec_2)
= \hm2\sqrt{\frac{S(p_1)S(p_2)}{S(k)}}
\left[\kvec\cdot\pvec_1\tilde X(p_1)+\kvec\cdot\pvec_2\tilde X(p_2)
-k^2\tilde X_3(\kvec,\pvec_1,\pvec_2)\right]
\label{eq:V3}
\end{equation}
where $\tilde X(p) = 1-1/S(p)$ is the ``direct correlation function''
and $\tilde X_3(\kvec,\pvec_1,\pvec_2)$ is the irreducible part
of the three-body distribution function, see
appendix \ref{app:X3}.

The lowest excitation branch is obtained by solving
\begin{equation}
\varepsilon_0(k) =\Re\, \Sigma(k,\varepsilon_0(k))\,.
\label{eq:disp}
\end{equation}
When consistent approximations are used, the solution of
Eq. (\ref{eq:disp}) is identical to what was obtained by CBF
perturbation theory.

Eqs. (\ref{eq:cbf}), (\ref{eq:sigmacbf}) and (\ref{eq:V3}) give the
correct physics up to and somewhat beyond the roton minimum, the
solution of Eqs. (\ref{eq:sigmacbf}), (\ref{eq:disp}) bridges about
about 80 percent of the discrepancy between the Feynman spectrum
$\varepsilon_{\rm F}(k)$ and the experiment.  The most prominent shortcoming
of the approximation is that it misses the energy of the plateau.  The
reason for this shortcoming is that the energy denominator in the
self-energy (\ref{eq:sigmacbf}) contains the Feynman energies.

\begin{figure}
    \centering
     \subfigure[]
     {\includegraphics[height=0.25\textwidth]{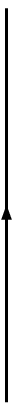}}\hspace{5em}
      \subfigure[]
     {\includegraphics[height=0.25\textwidth]{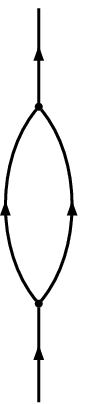}}\hspace{5em}
      \subfigure[]
     {\includegraphics[height=0.25\textwidth]{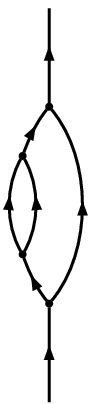}}\hspace{5em}
    \subfigure[]
    {\includegraphics[height=0.25\textwidth]{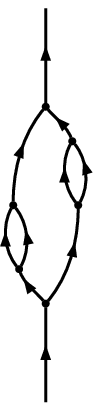}}
    \caption{Leading-order Feynman diagrams for the dynamic response
      function. (a) represents a single Feynman density wave, (b)
      shows the splitting into and recombination of two intermediate
      waves as described by pair fluctuations, whereas (c) and (d) are
      three-phonon excitations indicating the beginning of the
      self-consistent summation of (b). Processes of more complicated
      structure like one-to-three transitions have been neglected in
      the present calculation.}
\label{fig:diagrams}
\end{figure}

There are several ways to improve upon this: Brillouin-Wigner
perturbation theory has been worked out by Lee and Lee \cite{LeeLee}
up to fourth order from which the general scheme can be seen. Some
low-order processes contributing to the self-consistent self-energy
are shown in Fig.~\ref{fig:diagrams}.

Unfortunately that work did not utilize the fact that the ground state
should be optimized and, therefore, obtained also spurious diagrams.
The most complete derivation within the equations of motion scheme
includes time-dependent triplet correlations
\cite{eomIII,He4ModeMode}. The theory reproduces, for the lowest mode,
the first diagrams of CBF perturbation theory. The expected result is
that the self-energy in Eq. (\ref{eq:sigmacbf}) should be replaced by
the self-consistent form
\begin{equation}
\varepsilon_{\rm F}(p_1)+\varepsilon_{\rm F}(p_2)
\Longrightarrow
\Sigma(p_1,\hbar\omega-\varepsilon_{\rm F}(p_2))
+\Sigma(p_2,\hbar\omega-\varepsilon_{\rm F}(p_1))
\label{eq:sigmaself}\,,
\end{equation}
which leads to quantitative agreement between the theoretical excitation
spectrum and the experimental phonon-roton spectrum. It still contains
the Feynman energy as argument which should also be calculated
self-consistently. We have simplified this part of the calculation by
using the calculated phonon-roton spectrum in the energy arguments of
the self-energy. This provides a slight improvement of the description
in the momentum regime of the plateau.

\section{Dynamic structure function of two-dimensional $^4$He}

The only quantity needed for the calculation of the self-energy is the
ground state distribution function $g(r)$ and/or the static structure
function $S(k)$. These quantities have been calculated in the past and
are available in pedagogical and review-type literature, see
Refs.~\onlinecite{JordiQFSBook,MikkoQFSBook}. We have here used the HNC-EL
method including four and five-body elementary diagrams and triplet
correlation functions as described in Ref.~\onlinecite{EKthree}.

\begin{figure}
\centering
     {\includegraphics[width=0.33\textwidth,angle=-90]{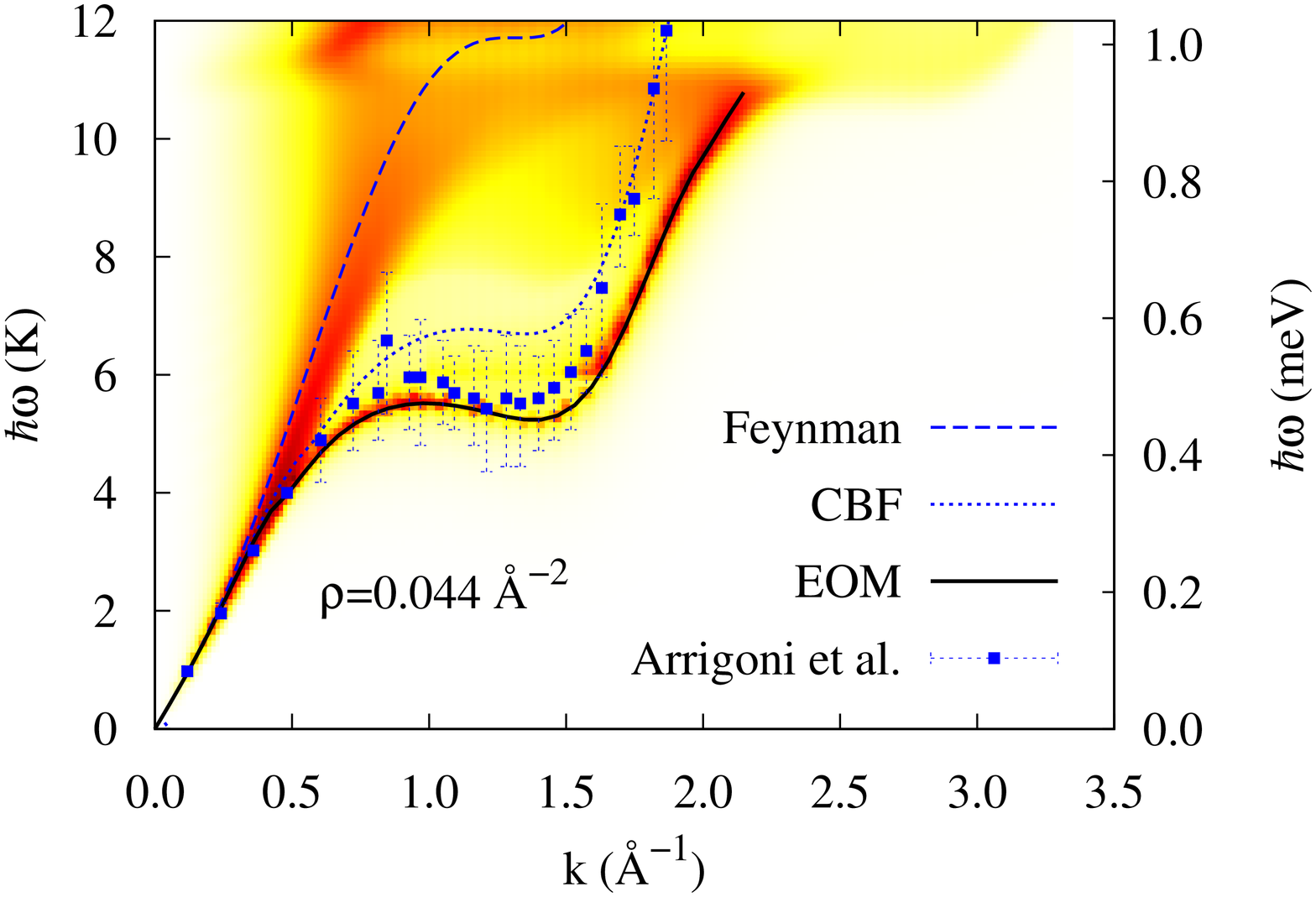}
\hspace{2em}
     \includegraphics[width=0.33\textwidth,angle=-90]{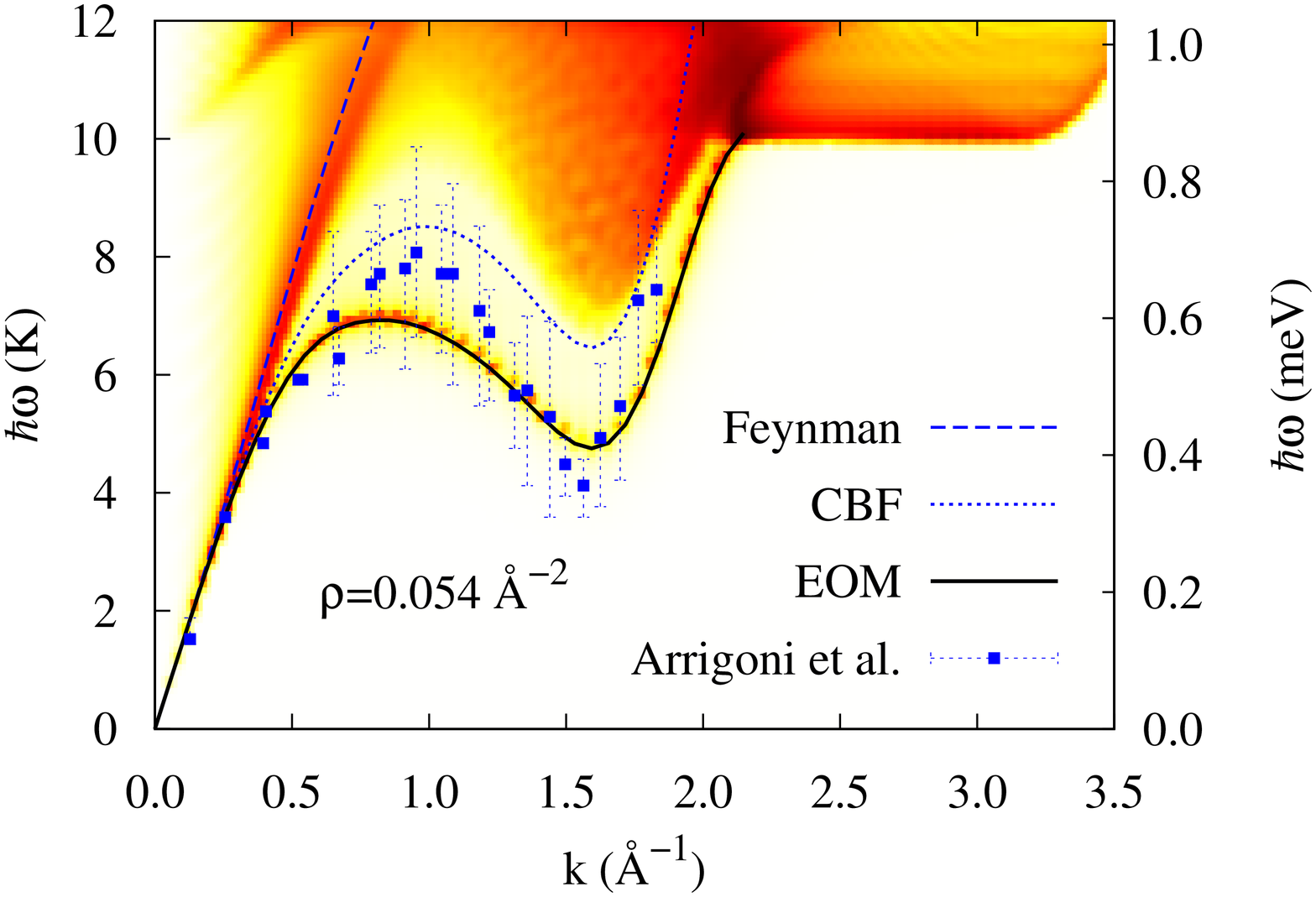}}\\
\centering
    {\includegraphics[width=0.33\textwidth,angle=-90]{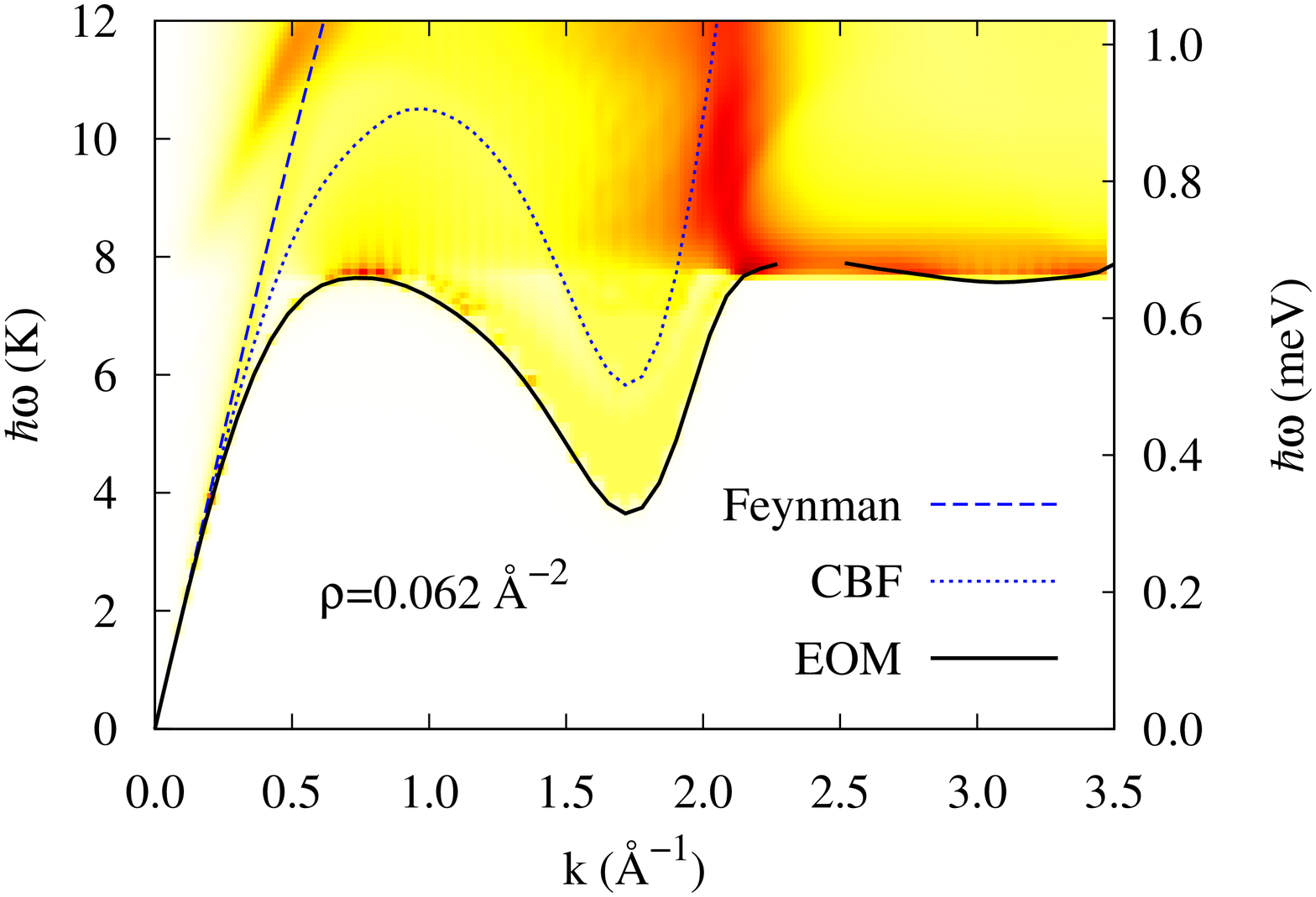}
\hspace{2em}
      \includegraphics[width=0.33\textwidth,angle=-90]{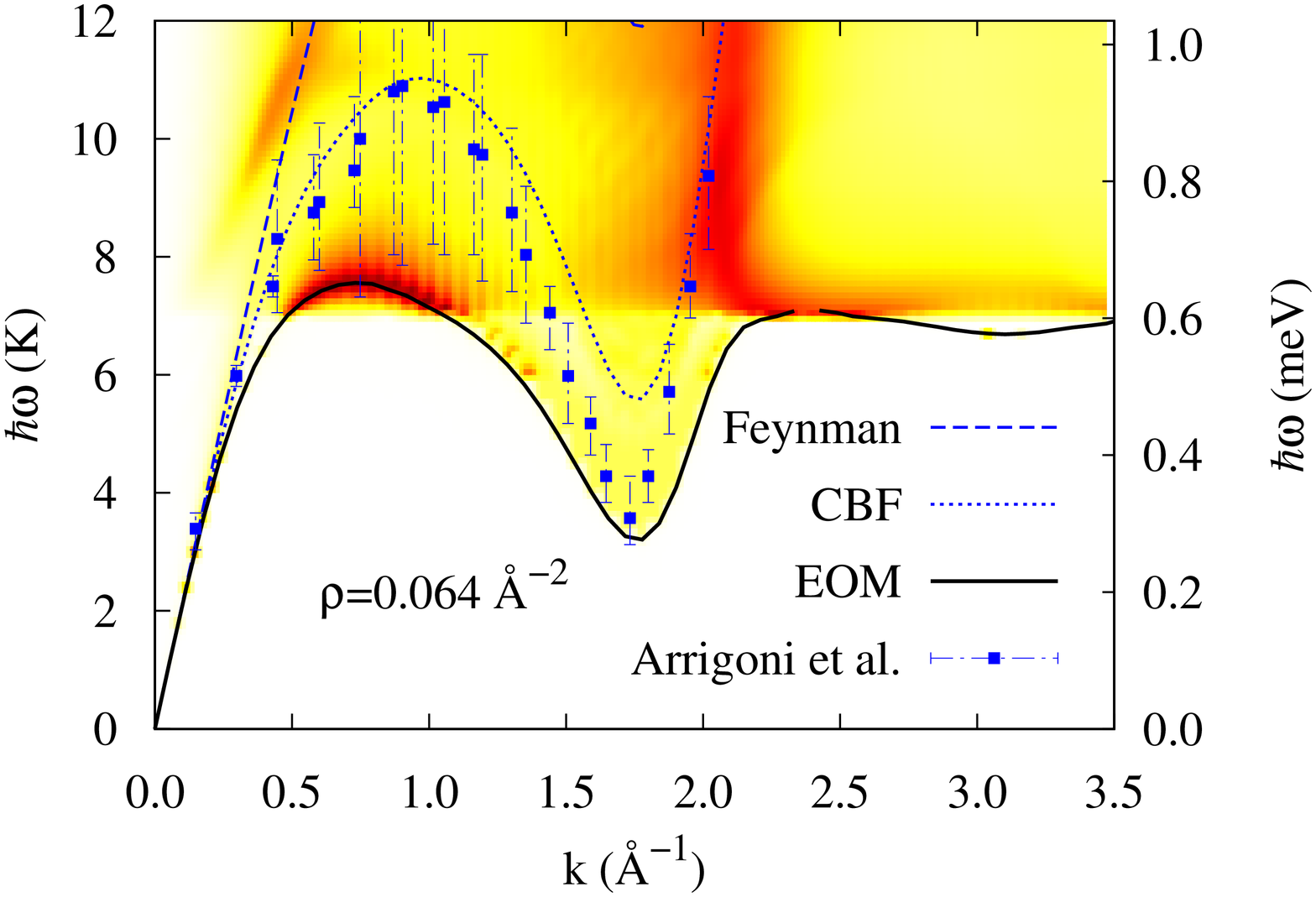}}
    \caption{(Color online) The figure shows contour plots of the
      dynamic structure function for a sequence of densities as shown
      in the legends. The colors have been chosen to highlight the
      prominent features, darker colors correspond to higher values of
      $S(k,\hbar\omega)$. The most striking observations are
      the appearance of a ``ghost phonon'' at low densities,
      and the presence of a secondary roton at
      high densities.  For
      comparison we also show the Feynman spectrum, the spectrum
      obtained within CBF-BW perturbation theory, and the simulation
      data of Ref.~\onlinecite{Arrigoni2013}.}
\label{fig:skwplots}
\end{figure}

We have calculated the dynamic structure function in the regime
between the equilibrium density of the system of $\rho =
0.042\,$\AA$^{-2}$ and the solidification density of $\rho =
0.064\,$\AA$^{-2}$ \cite{JordiQFSBook} in steps of $\Delta\rho =
0.002\,$\AA$^{-2}$.  Compared to earlier work we have used an improved
method for calculating the three-body vertex ${V}^{(3)}(\kvec;
\pvec_1,\pvec_2)$ as described in appendix \ref{app:X3}. This leads to
a slight lowering of the roton minimum by about $0.3\ldots 0.5$ K
depending on density but to no qualitative changes.  An overview of
our results for the dynamic structure function is shown in
Figs. \ref{fig:skwplots} for four densities. In these figures, we also
compare with the simulation data, including error bars, of
Ref.~\onlinecite{Arrigoni2013}.

\begin{figure}
\centerline{\includegraphics[width=0.50\textwidth,angle=-90]{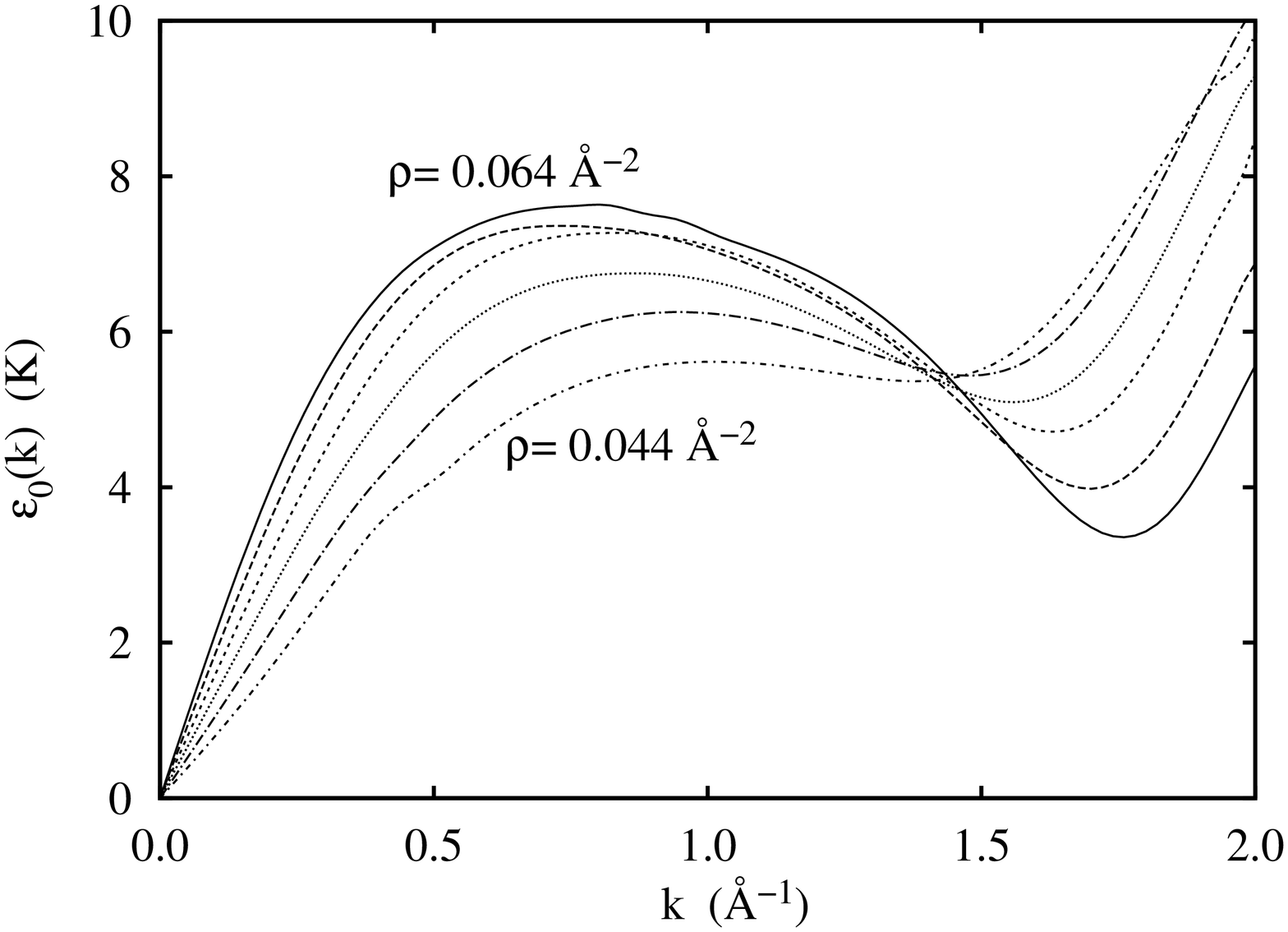}}
\caption{The figure shows the phonon-roton spectrum for the
densities $\rho=0.044\,$\AA$^{-2}$, $0.048\,$\AA$^{-2}$,\dots
 $0.064\,$\AA$^{-2}$, the curve with the lowest roton and the
highest maxon corresponds to the highest density. Note that the
dispersion for $\rho=0.044\,$\AA$^{-2}$ is anomalous, we have
in this case drawn the peak of $S(k,\hbar\omega)$. For a comparison
with available  simulation data, see Figs. \ref{fig:skwplots}.
\label{fig:ezs2d}}
\end{figure}

\begin{figure}
\centerline{\includegraphics[width=0.60\textwidth,angle=-90]{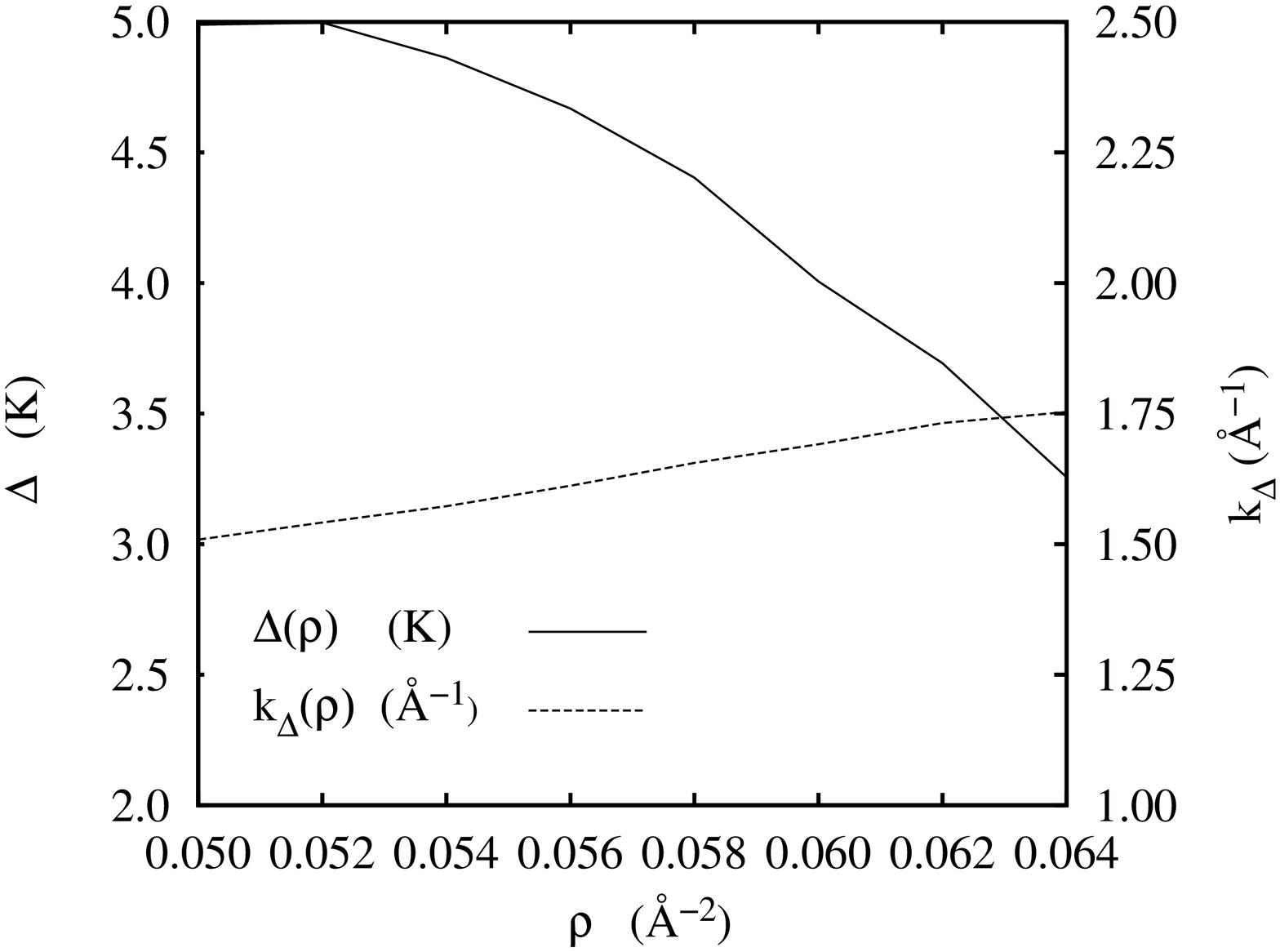}}
\caption{The figure shows the roton energy $\Delta$ (left scale) and the
roton wave number $k_\Delta$ (right scale) as a function of density
in the density regime $0.050\,$\AA$^{-2} \le \rho \le 0.064\,$\AA$^{-2}$.
For a comparison
with available  simulation data, see Figs. \ref{fig:skwplots}.
\label{fig:Delta}}
\end{figure}

Conventionally, one looks at the phonon-roton spectrum as the main
feature of the excitations in the helium liquids.  The phonon-roton
spectra are shown, as a function of density, in Fig. \ref{fig:ezs2d}.
These spectra display, apart from an energy and momentum scale which
is distinctly different from the three-dimensional case, similarities to
the 3D spectra: with increasing density, the roton energy is lowered,
and the roton wave number becomes larger. The roton is normally
described by the parameters roton energy $\Delta$, roton wave
number $k_\Delta$ and roton ''effective mass'' $\mu$,
\begin{equation}
\varepsilon_0(k) = \Delta + \frac{\hbar^2}{2\mu}(k-k_\Delta)^2\,.
\label{eq:efit}
\end{equation}
Fig. \ref{fig:Delta} shows the density dependence of the roton energy
and wave number. For that purpose, we have fitted the spectra in a
regime of $k = k_\Delta \pm 0.15\,$\AA$^{-1}$ by the form
Eq.~(\ref{eq:efit}). The values of $\Delta$ and $k_\Delta$ are somewhat
sensitive to the choice of the momentum range used for the the fit,
especially at the lower densities where the roton minimum is not very
pronounced.  Because of this we refrain from showing a comparison with
the simulation data in Fig.~\ref{fig:Delta}, Figs. \ref{fig:skwplots}
contain the same information but include error bars and are more
informative.  These pieces of information are the standard quantities
that characterize the phonon-roton spectrum.

Let us now focus on those features of the dynamic structure function
where the 2D case differs visibly from the 3D system:
\begin{itemize}
\item {} It was already noted in Ref.~\onlinecite{JordiQFSBook} that the
  speed of sound is low compared to the same quantity in 3D. The
  consequence is a strong anomalous dispersion which has, in turn, the
  consequence that long wavelength phonons can decay up to a density
  of about $0.050\,$\AA$^{-2}$.
\item{} Similar to the 3D case we notice at low to moderate densities
  a feature which was tentatively called ``ghost phonon''
  \cite{He4ModeMode}.  In contrast to the 3D system, where the ghost
  phonon disappears rapidly with increasing density, the feature is
  very pronounced even at a density of $\rho = 0.054\,$\AA$^{-2}$.
\item{} At very high densities, slightly below the liquid-solid phase
  transition, we see a mode that is clearly separated from the plateau.
  The plateau itself is a threshold above which an induced density
  fluctuation of wave vector $\kvec$ and frequency $\omega$ can decay,
  under energy and momentum conservation, into two rotons. 
  This condition can be satisfied for \textit{all} momenta below twice the roton momentum. At high densities a signature of the resulting discontinuity in the imaginary part of the self-energy is visible not only beyond but also in the roton and even maxon regions.
 \end{itemize}

\begin{figure}
\centerline{\includegraphics[width=0.33\textwidth,angle=-90]{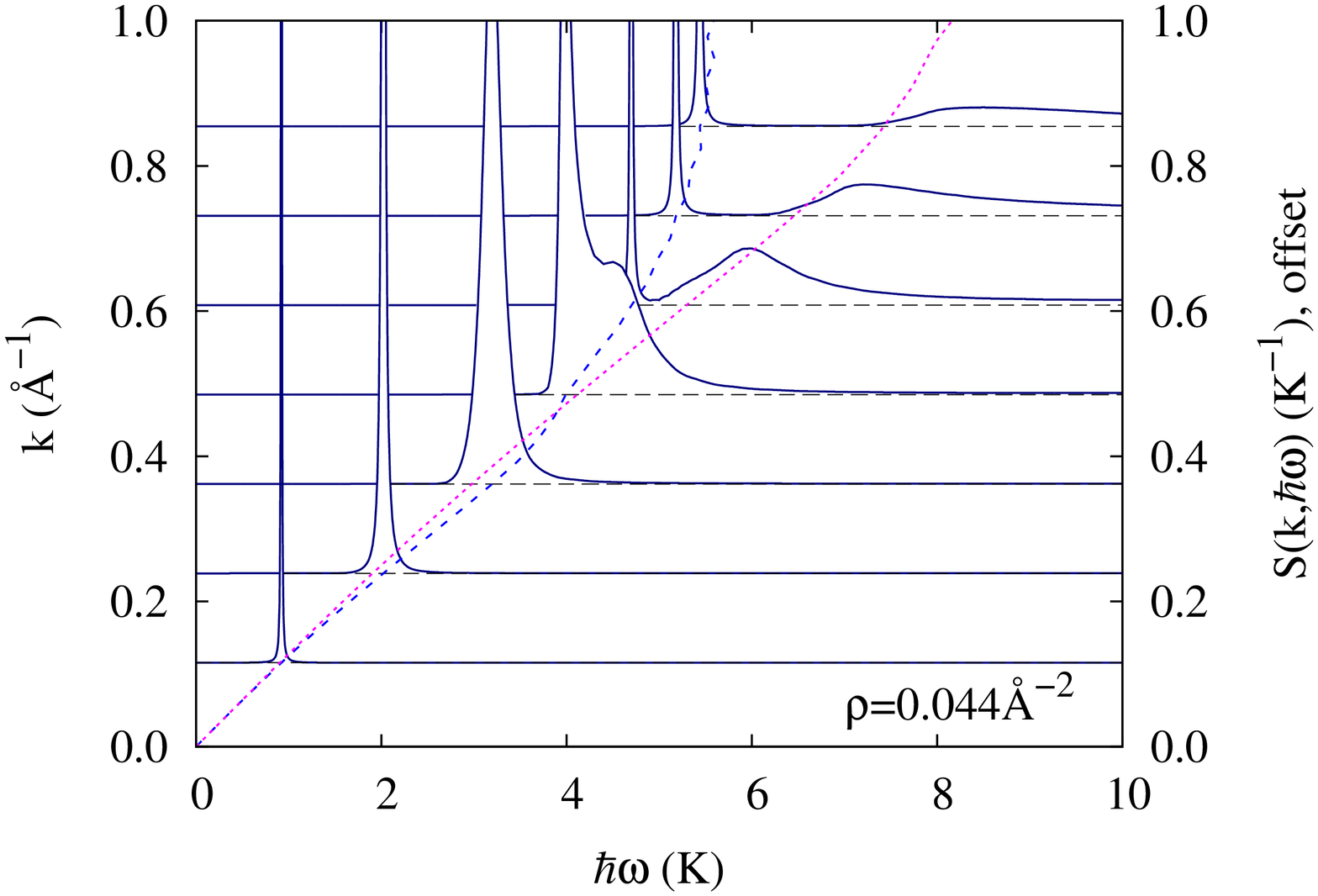}
\hspace{2em}
\includegraphics[width=0.33\textwidth,angle=-90]{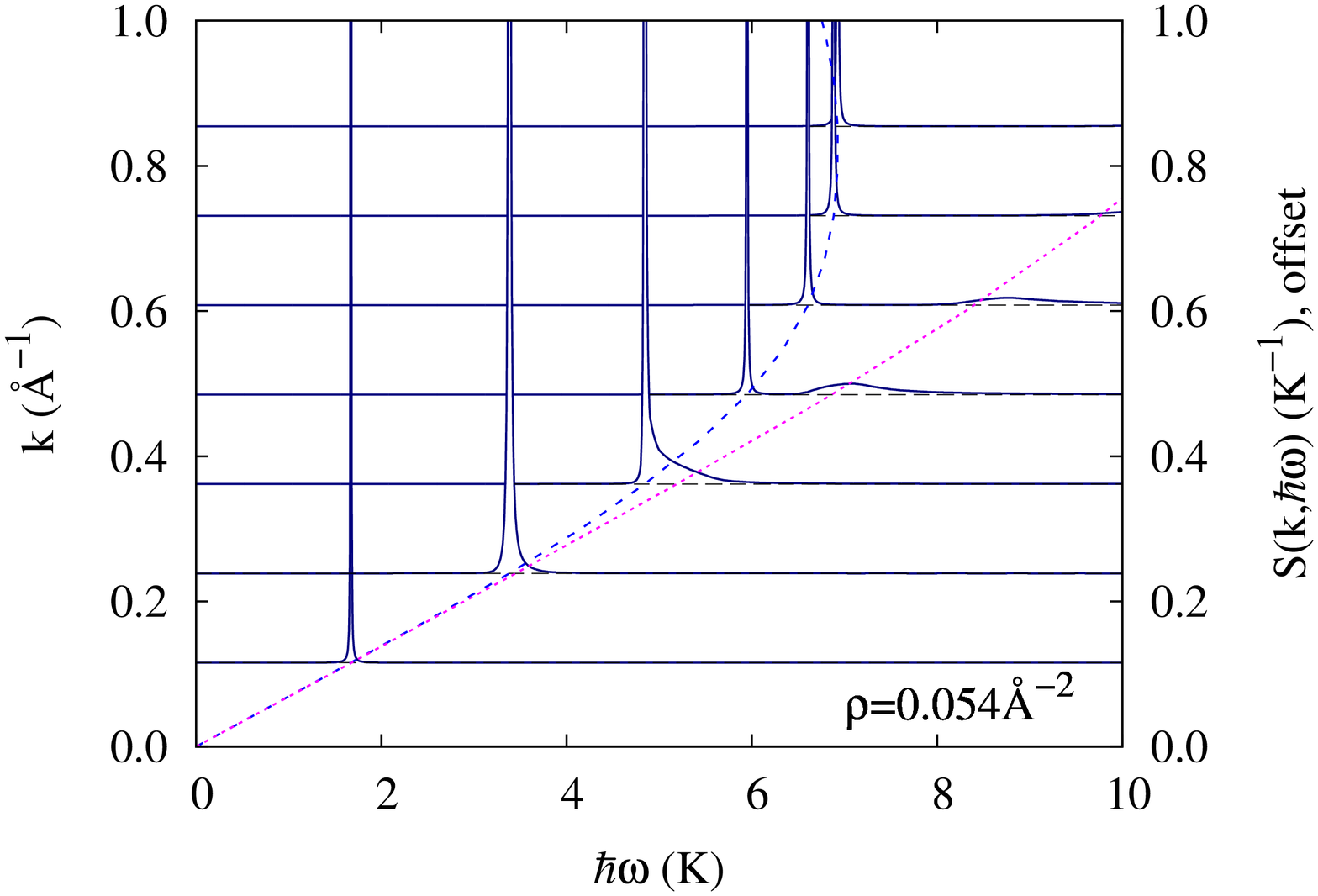}}
\caption{The figures show cuts of $S(k,\hbar\omega)$ at long wavelengths
  at the densities $\rho = 0.044\,$\AA$^{-2}$ (left pane) and $\rho =
  0.054\,$\AA$^{-2}$ (right pane). The long dashed line is the phonon
  dispersion relation, and the short-dashed line is the curve
  $2\varepsilon_0(k/2)$. At higher density, the phonon becomes sharper but
  the ghost phonon is still visible.
\label{fig:ghosts}}
\end{figure}

 We have noted above that anomalous dispersion persists well beyond
 equilibrium density. This leads to the damping of long-wavelength
 phonons.  Figs. \ref{fig:ghosts} show cuts of $S(k,\hbar\omega)$ at
 long wavelengths. At the first glance, it appears that the phonon
 broadens at a wave number of $k\approx 0.38\,$\AA$^{-1}$. Closer
 inspection reveals, however, that a second, broad feature splits off
 the phonon and becomes an isolated feature above $k\approx
 0.6\,$\AA$^{-1}$. Eventually the feature dissolves around $k\approx
 1.0\,$\AA$^{-1}$. The effect is also seen quite clearly in the two
 contour plots corresponding to the densities $0.044\,$\AA$^{-2}$ and
 $0.054\,$\AA$^{-2}$ shown in Figs. \ref{fig:skwplots}. On the other
 hand, the broadening that should occur, due to anomalous dispersion,
 up to wave numbers of about 0.4\,\AA$^{-1}$, is hardly visible.

 The feature can be explained by examining the analytic structure of
 the self-energy in 2D. Specifically, we will show in appendix
 \ref{app:anomalous} that the imaginary part of the self-energy has,
 in the limit $\hbar\omega\rightarrow 2\varepsilon_0(k/2)$, a discontinuity of
 the form
\begin{eqnarray}
\Im\Sigma(k,\hbar\omega&\rightarrow& 2\varepsilon_0(k/2)) \sim \nonumber\\
&-&\sqrt{\frac{k}{2\varepsilon'_0(k/2)\varepsilon_0''(k/2)}}
\theta\Big(\mathrm{sign}\big(\varepsilon_0''(k/2)\big)\big(\hbar\omega-2\varepsilon_0(k/2) \big)\Big),
\label{eq:Sigmalimit}
\end{eqnarray}
which implies a logarithmic singularity of
$\Re\Sigma(k,\hbar\omega\rightarrow 2\varepsilon_0(k/2))$.
Eq. (\ref{eq:Sigmalimit}) is normally derived for the purpose of
estimating the lifetime of phonons in the regime of
anomalous dispersion \cite{bosegas}. However, it is also valid for
normal dispersion $\varepsilon_0''(k/2) < 0$ as long as $|k
\varepsilon_0''(k/2)| \ll \varepsilon_0'(k/2)$, {\em i.e.\/} one should see the
signature of the step function of the imaginary part of the
self--energy up to about twice the wave number for which the
dispersion relation $\varepsilon_0(k)$ is, to a good approximation, linear.
This is exactly the regime where the ghost phonon is seen in
Figs. \ref{fig:skwplots}. We also note that the effect is stronger in
2D than in 3D because there the logarithmic singularity
$\ln\left(2\varepsilon_0(k/2)-\hbar\omega\right)$ in the real part of the
self-energy giving rise to Eq. (\ref{eq:Sigmalimit}) is replaced by
$\sqrt{2\varepsilon_0(k/2)-\hbar\omega}$ \cite{bosegas}.

\begin{figure}
\centerline{\includegraphics[width=0.40\textwidth,angle=-90]{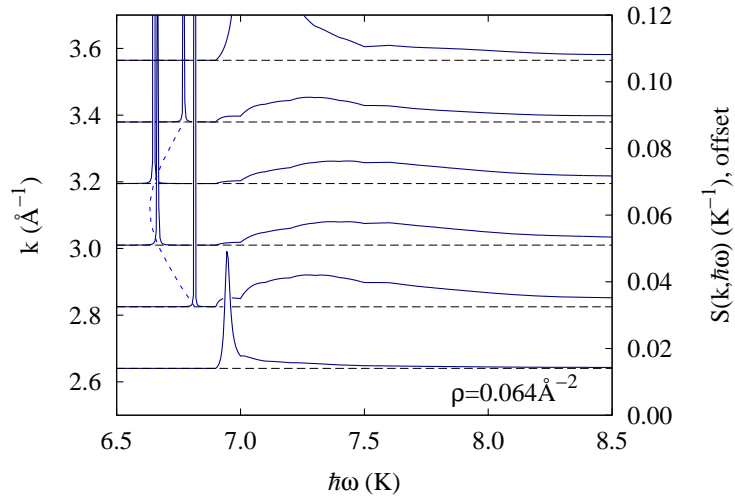}
}
\caption{The figure shows cuts of $S(k,\hbar\omega)$ in the regime
  of the Pitaevskii plateau for the density $\rho =
  0.064\,$\AA$^{-2}$ for a sequence of momenta
  $ 2.6\,$\AA$^{-1}\,\le k\,\le 3.6 $\AA$^{-1}$.
\label{fig:stack2}}
\end{figure}

A second striking feature is the appearance of a sharp mode below the
plateau.  We stress the difference: normally, the plateau is a
threshold above which a wave of energy/momentum $(\hbar\omega, k)$ can
decay into two rotons. This has the consequence that the imaginary
part of the self-energy $\Sigma(k,\hbar\omega)$ is a step function and
the real part has a logarithmic singularity \cite{Pitaevskii2Roton}. A
collective mode is, on the other hand, characterized by a singularity
of the $S(k,\hbar\omega)$. Figs. \ref{fig:skwplots} show, for the two
highest densities, the appearance of a sharp discrete mode {\em
  below\/} the plateau. A close-up of the situation is shown in
Fig. \ref{fig:stack2}: Clearly the plateau starts at the same energy
for all momenta. At a wave number of $k\approx 2.6\,$\AA$^{-1}$, the
collective mode is still merged into the continuum. With increasing
wave number, we see, however, a clearly distinguishable mode about 0.3
K below the plateau.

\section{Discussion}
\label{sec:Discussion}

We have already made the essential points of our findings in the
discussion of our results. Evidently, the difference between two and
three dimensions has quite visible effects on $S(k,\hbar\omega)$, as
mentioned above.

Our findings about a secondary roton should shed some light on the
discussion of the nature of the roton minimum. It has been argued
\cite{FeynmanPLTP} that the roton is the ``ghost of a vanished vortex
ring'' or \cite{NozSolid,nozieres2006} ``ghost of a Bragg Spot'' due
to the imminent liquid-solid phase transition.  In this density region
two-dimensional $^4$He already shows a strong signature of the
triangular lattice into which it eventually
freezes~\cite{whitlock1988monte,halinen2000excitations,apaja2000liquid}
and can exhibit a so called hexatic phase~\cite{apaja2008structure}.

If the Bragg spot interpretation of the roton is correct, one should
perhaps expect a second one and the ratio of the absolute values of
the corresponding wave vectors should roughly satisfy $k_2/k_1=
2\cos(\pi/6)\approx 1.73$ because of the triangular lattice of the
solid phase.

Our results may indeed be interpreted as an indication that this is
the case. It is certainly worth investigating this issue further along
the line of angular-dependent excitations \cite{SmithSolid}. A similar
effect has been seen in cold dipolar gases
\cite{PhysRevLett.109.235307} and the relationship is worth examining.

Finally a word about the comparison with simulation data
\cite{Arrigoni2013}. Overall, the agreement appears satisfactory, most
of our results are within the error bars of that calculation. The most
visible discrepancy is seen at the highest density of
$\rho=0.064\,$\AA$^{-2}$. At this density, the maxon energy is below
twice the roton energy and the modes in this $(\hbar\omega,k)$ region
can decay.  One would expect more strength at the decay threshold of
$2\Delta$ as shown by our results whereas the Monte Carlo data
indicate -- despite large error bars -- that the decay strength lies
at higher energies.  This point deserves further investigation, it
might also explain why the maxon energies at $\rho=0.054\,$\AA$^{-2}$
differ more than expected. Otherwise the agreement is quite good,
evidently the strength shown at and above $k = 2\,$\AA$^{-1}$ follows
indeed the kinetic energy branch in both calculations, whereas the
plateau region has relatively little strength.

\appendix

\section{Long-wavelength dispersion in 2D}
\label{app:anomalous}
In this appendix, we study the analytic structure of the self-energy
as a function of an external energy $\hbar\omega$ in the limit
${\hbar\omega - 2 \varepsilon_0(k/2)\rightarrow 0}$.  We assume that the
solution $\varepsilon_0(k)$ of the implicit equation (\ref{eq:disp})
has a negligible imaginary part.

We look for processes where a state of wave vector ${\bf k}$ decays
into two phonons of wave vectors ${\bf p}_1$ and ${\bf
  p}_2$. In general one expects, for long wavelengths, a phonon
dispersion relation of the form
\begin{equation}
\varepsilon_0(k) = \hbar c k + c_3 k^3
\end{equation}
where $c$ is the speed of sound. In fact, it is easily shown
that Eq. (\ref{eq:disp}) leads to such a dispersion relation.

The calculation is best carried out in relative and center of mass
momenta, {\em i.e.\/} we set
\[\pvec_1 = \qvec- \frac{1}{2}\kvec\qquad\pvec_2 = -\qvec- \frac{1}{2}\kvec.\]
Then, it is clear that
\begin{equation}
\varepsilon_0(|\kvec/2+\qvec|) +\varepsilon_0(|{\bf k}/2-\qvec|)
\end{equation}
has, for all angles $\cos\theta \equiv x \equiv \hat\qvec\cdot\hat\kvec$,
a relative extremum at $q=0$. Expanding the energy denominator as
\begin{equation}
\varepsilon_0(p_1)+\varepsilon_0(p_2) = 2\varepsilon_0(k/2) +
\left[\frac{2\varepsilon_0'(k/2)}{k}(1-x^2) +
\varepsilon_0''(k/2)x^2\right]q^2 +{\cal O}(q^3)\,,
\end{equation}
we see that the the value $2\varepsilon_0(k/2)$ is, at $x=1$, a relative minimum
if $c_3 > 0$ (anomalous dispersion) and relative maximum for $c_3 < 0$
(normal dispersion).
For further reference, abbreviate
\[\varepsilon_0'\equiv\varepsilon_0'(k/2)\,,\qquad \varepsilon_0''\equiv
\varepsilon_0''(k/2)\,.\]

We do the calculation first for the case of anomalous dispersion.
The three-body coupling matrix element assumes
a finite value as $\qvec\rightarrow 0$, we therefore need
to include only the leading term
\[
V_3\left({\bf k};-\frac{1}{2}{\bf k}+\qvec,-\frac{1}{2}{\bf k}-\qvec\right)
\approx V_3\left({\bf k};-\frac{1}{2}{\bf k},-\frac{1}{2}{\bf k}\right)
\]
Then
\begin{eqnarray}
  \Im\Sigma(k,\hbar\omega) &\approx&
  \frac{ \left|V^{(3)}({\bf k};-\frac{1}{2}{\bf k},-\frac{1}{2}{\bf k})\right|^2}
  {2(2\pi)^2\rho}
  \Im\int \frac{d^2q}
  {\hbar\omega-\varepsilon_0(p_1)-\varepsilon_0(p_2)+\I\eta}\nonumber\\
  &\approx&
  \frac{ \left|V^{(3)}({\bf k};-\frac{1}{2}{\bf k},-\frac{1}{2}{\bf k})\right|^2}
  {2(2\pi)^2\rho}
  \Im \int \frac{d^2q}
  {e_0(q)+(e_1(q)-e_0(q))cos^2\theta+\I\eta}\\
  &=&
  -\frac{ \left|V^{(3)}({\bf k};-\frac{1}{2}{\bf k},-\frac{1}{2}{\bf k})\right|^2}
  {4\pi\rho}
  \int_{q_-}^{q_+} \frac{q dq}
  {\sqrt{-e_0(q)e_1(q)}}
\end{eqnarray}
where $e_0(q)\equiv\hbar\omega-2\varepsilon_0(k/2) -\frac{2\varepsilon_0'}{k}q^2$ and
$e_1(q)\equiv\hbar\omega-2\varepsilon_0(k/2) -\varepsilon_0''q^2$ are the values of the
energy denominator at $x=0$ and $x=1$. The integral is imaginary if
the denoninator changes its sign for ${0\le \cos^2\theta\le 1}$.  Since
per assumption $\varepsilon_0''\ll 2\varepsilon_0'/k$ we have always $e_0(q)
< e_1(q)$, therefore we need $\Delta E\equiv \hbar\omega - 2\varepsilon_0(k/2)>0$
to have an imaginary part. Then, because of $\varepsilon_0'' >0$, the imaginary part
is picked up for $q_- < q < q_+$, where
\begin{equation}
q_-=\sqrt{\frac{k\Delta E}{2\varepsilon_0'}} \qquad
q_+=\sqrt{\frac{\Delta E}{\varepsilon_0''}}\,,
\end{equation} 
and, hence
\begin{eqnarray}
\Im\Sigma(k,\hbar\omega) &=&
-\frac{ \left|V^{(3)}({\bf k};-\frac{1}{2}{\bf k},-\frac{1}{2}{\bf k})\right|^2}
{4\pi\rho}
        \int_{q_-}^{q_+} \frac{q dq}
{\sqrt{-e_0(q)e_1(q)}}\theta(\Delta E)\nonumber\\
&=&-\sqrt{\frac{k}{2\varepsilon'_0\varepsilon_0''}}
\frac{\left|V_3({\bf k};-\frac{1}{2}{\bf k},
-\frac{1}{2}{\bf k})\right|^2}{8\rho}
\theta(\Delta E)\,.
\label{eq:Imsigma2da}
\end{eqnarray}

For $\varepsilon_0''<0$, there is no upper limit of the
integration range of the internal momentum, but the integral converges
because the three-phonon matrix element goes to zero
for large momentum transfers, and the actual value depends
on the details of the interaction. However, we are interested
only in the non-analytic behavior for
$\Delta E\rightarrow 0$.
To calculate this behavior, subtract and add the matrix element
at the position where the denominator has a second order node,
{\em i.e.} we write
\begin{equation}
\left|V_3({\bf k};{\bf q}-\frac{1}{2}{\bf k},
-{\bf q}-\frac{1}{2}{\bf k})\right|^2
= \left|V_3({\bf k};-\frac{1}{2}{\bf k}
-\frac{1}{2}{\bf k})\right|^2  + \Delta V({\bf k},{\bf q}).
\end{equation}
$\Delta V({\bf k},{\bf q})$ still contributes to the imaginary
part, but not to the non-analytic behavior.
We must now distinguish between $\Delta E > 0$
and $\Delta E < 0$.
For the former case we have 
\begin{eqnarray}
\Im \Sigma(k,\hbar\omega)
&\approx&
-\frac{\left|V_3({\bf k};-\frac{1}{2}{\bf k},
-\frac{1}{2}{\bf k})\right|^2}{4\pi\rho}
\int_{q_-}^{\ldots}\frac{q dq}
{\sqrt{-e_0(x)e_1(x)}}.
\nonumber
\end{eqnarray}
The momentum integral does not converge, but this
is artificial because we have factored out the interaction since
we are only interested in the behavior due to the
square-root singularity at $q_-$. Therefore, write
for $\Delta E>0$
\begin{eqnarray}
&&\int_{q_-}^{\dots}\frac{qdq}{\sqrt{(\frac{2\varepsilon_0'}{k}q^2-\Delta E) 
(\Delta E +|\varepsilon_0''|q^2)}}\nonumber\\
&=&\frac{1}{\Delta E}
\int_{q_-}^{\dots}\frac{qdq}{\sqrt{(\frac{q^2}{q_-^2}-1) 
(1 +\frac{|\varepsilon_0''|}{\Delta E}q^2)}}
\nonumber\\
&=&\frac{q_-^2}{\Delta E}\int_1^{\dots}\frac{x dx}{\sqrt{(x^2-1) 
(1 +\frac{k|\varepsilon_0''|}{2\varepsilon_0'}x^2)}}
= \frac{k}{2\varepsilon_0'}\times({\rm a\ number}).
\end{eqnarray}
We can ignore the term $k\varepsilon_0''/2\varepsilon_0'$ because, by
assumption,
$|\varepsilon_0''|\ll\varepsilon_0'/k$. The integral is then just a numerical
value.

For $\Delta E<0$ we get
\begin{eqnarray}
&&\int_{q_+}^{\dots}\frac{dq}{\sqrt{(|\Delta E| + \frac{2\varepsilon_0'}{k}q^2)
(|\varepsilon_0''|q^2-|\Delta E|)}}\nonumber\\
&=&
\frac{1}{|\Delta E|}
\int_{q_+}^{\dots}\frac{qdq}{\sqrt{(1+\frac{2\varepsilon_0'}{k|\Delta E|}q^2)
(\frac{|\varepsilon_0''|}{|\Delta E|}q^2-1)}}
\nonumber\\
&=&\frac{q_+^2}{|\Delta E|}\int_1^{\dots}\frac{x dx}{\sqrt{(x^2-1) 
(1 +\frac{2\varepsilon_0'}{k|\varepsilon_0''|}x^2)}}
= \sqrt{\frac{k}{2\varepsilon_0'\varepsilon_0''}}\times({\rm a\ number}).
\end{eqnarray}
Here, the term $\frac{2\varepsilon_0'}{k|\varepsilon_0''|}$
dominates in the denominator. Since, by assumption, $k|\varepsilon_0''|
\ll \varepsilon_0'$, the imaginary part has a discontinuity of the order
of $\sqrt{\frac{k}{2\varepsilon_0'\varepsilon_0''}}$ at $\Delta E = 0$.

\section{Three-Body Vertex}
\label{app:X3}
\begin{figure}
\centerline{\includegraphics[width=0.9\textwidth]{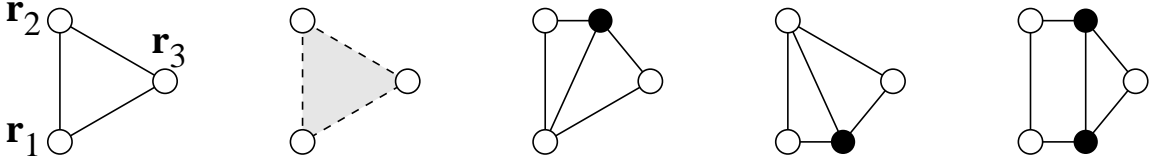}}
\caption{The figure shows the leading order diagrams contributing to
  the irreducible three-body vertex
  $X_3(\rvec_1,\rvec_2,\rvec_3)$. The usual diagrammatic conventions
  apply: circles correspond to particle coordinates, filled circles
  imply a density factor and integration over the associated
  coordinate space. Solid lines represent correlation factors
  $h(\rvec_i,\rvec_j)=g(\rvec_i,\rvec_j)-1$ and the shaded triangle
  represents a three-body function $u_3(\rvec_1,\rvec_2,\rvec_3)$.}
\label{fig:rhs}
\end{figure}

Normally, the three-body vertex (\ref{eq:V3}) is calculated in
convolution approximation. An improvement can be achieved by
summing a set of three-body diagrams contributing to
$\tilde X_3(\kvec,\pvec,\qvec)$, which corresponds topologically
to the hypernetted chain (HNC) summation. The first few diagrams
are shown in Fig \ref{fig:rhs}.

The equations to be solved are best written in momentum space and relative
and center of mass momenta, {\em i.e.}
\begin{equation}
\tilde X(\pvec_1,\pvec_2,\pvec_3) \equiv
\tilde  X(\qvec/2+\kvec,\qvec/2-\kvec,\qvec)\equiv \tilde X_{\qvec}(\kvec)\,.
\end{equation}
 The integral equation to be solved is
\begin{eqnarray}
\tilde X_{\qvec}(\kvec) &=& \int\frac{ d^dp}{(2\pi)^d\rho}
\tilde h(\kvec-\pvec)\tilde N_{\qvec}(\pvec)\nonumber\\
\tilde N_{\qvec}(\kvec) &=&\tilde N^{(CA)}_{\qvec}(\kvec)
+ \tilde s_{\qvec}(\kvec)\delta\tilde X_{\qvec}(\kvec),
\end{eqnarray}
where $\tilde N_{\qvec}(\kvec)$ is the set of nodal diagrams, and
\[\tilde N^{(CA)}_{\qvec}(\kvec) = \tilde h(\frac{\qvec}{2}+\kvec)
\tilde h(\frac{\qvec}{2}-\kvec) +
\tilde u_3(\frac{\qvec}{2}+\kvec,\frac{\qvec}{2}-\kvec,\qvec)\]
is the convolution approximation for this quantity.
Also, we have abbreviated
\begin{equation}
\tilde s_{\qvec}(\kvec) = \left[S(|\pvec+\qvec/2|)S(|\pvec-\qvec/2|)-1\right]\,.
\end{equation}

The equations can be easily solved by expanding all functions
in terms of $k$, $q$, and the angle between the two vectors, {\em e.g.\/}
\[\tilde h(|\kvec_1-\kvec_2|) =\sum_{n=0}^\infty \tilde h_n(k_1,k_2)
\cos(n\,\phi_{12})\]

This gives us the three-body vertex in the form
\[\tilde X_{\qvec}(\pvec) = \sum_m\cos( m\phi) X_m(q,p)\,.\]

\begin{acknowledgements}
  We would like to thank C. E. Campbell, F. Gasparini and H. Godfrin
  for useful discussions.  This work was supported, in part, by the
  Austrian Science Fund FWF under project I602.  Additional support
  was provided by a grant from the Qatar National Research Fund
  \# NPRP NPRP 5 - 674 - 1 - 114.
\end{acknowledgements}

\bibliography{papers,morepapers}{}
\bibliographystyle{spphys}

\end{document}